\documentstyle[12pt]{article}
\def\title #1 {
   \headsep=0.6in
   \baselineskip=30pt
			\begin{center}
   {\titlebold #1}
   \end{center}
			\vskip .50in }
\def\author #1 {
   \baselineskip=30pt
   \begin{center}
   {\timeslarge #1}
   \end{center}
			\vskip .25in }
\def\address #1 {
   \baselineskip=24pt
   \begin{center}
   {\timesitalic #1}
   \end{center}
   \vskip 0.8in }

\renewcommand{\title}[1]{\large\bf \mbox{}\\ \mbox{}\\ \mbox{}\\ \mbox{}\\
     #1\bigskip\medskip\\}
\renewcommand{\author}[1]{\large #1\\ \smallskip}
\renewcommand{\address}[1]{{\narrower\normalsize\it #1\\}\bigskip}
\renewenvironment{abstract}{\narrower\small}{\par\normalsize\bigskip}

\renewcommand{\thefootnote}{\fnsymbol{footnote}}

\newcounter{num}

\newcommand{\be}{\begin{equation}}
\newcommand{\ee}{\end{equation}}
\newcommand{\ba}{\begin{eqnarray}}
\newcommand{\ea}{\end{eqnarray}}

\newcommand{\nonu}{\nonumber \\[2mm]}

\newcommand{\cd}{c^{\dagger}}

\newcommand{\up}{\uparrow}
\newcommand{\down}{\downarrow}
\newcommand{\si}{\sigma}

\newcommand{\vac}{| \, 0 \, \rangle}

\newcommand{\ferro}{|\hskip-3pt\down\cdots \down \rangle}
\newcommand{\ferroup}{|\hskip-3pt\up\cdots \up \rangle}
\newcommand{\sumnn}{\sum_{j=1}^L}
\newcommand{\grst}{|\,\psi_N\,\rangle}
\newcommand{\half}{{1\over 2}}

\newcommand{\pr}{Phys.\ Rev.\ }
\newcommand{\prl}{Phys.\ Rev.\ Lett.\ }

\renewcommand{\thefootnote}{\fnsymbol{footnote}}
\begin{document}
\begin{center}
\titlepage
\
\vskip-0.8truecm
\hfill ITP-SB-94-45
\vspace*{-0.3cm}
\title{$XXZ$ model as effective Hamiltonian for generalized
Hubbard models with broken $\eta$-symmetry}

\vskip1cm

\author{G.\ Albertini$^{a,b,}$\footnote{email:
{\tt albert@pib1.physik.uni-bonn.de}},
V.~E.\ Korepin$^{a,}$\footnote{email: {\tt korepin@insti.physics.sunysb.edu}}
and A.\ Schadschneider$^{a,}$\footnote{present address:
Institut f\"ur Theoretische Physik, Universit\"at zu K\"oln, 50937 K\"oln,
Germany (email: {\tt as@thp.uni-koeln.de})}}
\vspace*{0.2cm}
\address{$^a$ Institute for Theoretical Physics\\
SUNY at Stony Brook\\
Stony Brook, NY 11794-3840, USA}

\address{$^b$ Physikalisches Institut \\
Universit\"at Bonn\\
53115 Bonn, Germany}

\vskip0.2cm
\date{\today}
\end{center}
%\vspace*{0.2cm}
\centerline{\today}

\vskip0.7cm

%\maketitle
%\thispagestyle{empty}
\begin{abstract}
\normalsize
\noindent
We consider the limit of strong Coulomb attraction for generalized
Hubbard models with $\eta$-symmetry. In this limit these models are
equivalent to the ferromagnetic spin-1/2 Heisen\-berg quantum spin
chain.  In order to study the behaviour of the superconducting phase
in the electronic model under perturbations which break the
$\eta$-symmetry we investigate the ground state of the ferromagnetic
non-critical $XXZ$-chain in the sector with fixed magnetization. It
turns out to be a large bound state of $N$ magnons. We find that the
perturbations considered here lead to the disappearance of the
off-diagonal longe-range order.
\end{abstract}
\vspace*{0.3cm}
Short title:  $XXZ$ model as effective Hamiltonian for gen.~Hubbard\\
%\vspace*{0.3cm}

\newpage

\renewcommand{\thefootnote}{\arabic{footnote}}
\setcounter{footnote}{0}

There has been a lot of activity in the past few years in the investigation
of electronic models for superconductivity. Especially the so-called
$\eta$-pairing mechanism introduced by Yang \cite{YANG} has been
quite fruitful for the construction of models with superconducting
ground states (see e.g.~\cite{dBKS}). Although the ground state
of the Hubbard model is not of the $\eta$-pairing type \cite{YANG}
one can construct such models by adding nearest-neighbour interaction
terms. The first example was found in \cite{EKSa,EKSb}. In \cite{dBKS}
it has been shown that for a large class of generalized Hubbard models
the ground state in the limit $U\to -\infty$ is the simplest $\eta$-pairing
state $(\eta^\dagger)^{N}|0\rangle$ with $\eta^\dagger = \sum_j
c_{j\down}^\dagger c_{j\up}^\dagger$. For the supersymmetric Hubbard
model \cite{EKSa,EKSb} and a special case of the Hirsch model with
correlated hopping interaction \cite{as} the complete phase
diagram in arbitrary dimensions has been found. It
shows two phases with $\eta$-pairing ground states. Since these
states have ODLRO they are also superconducting \cite{ODLRO}.

The main aim of this letter is to clarify the effect of
$\eta$-symmetry-breaking perturbations on the superconducting ground
state. In order to keep the problem as simple as possible we restrict
ourselves to the one-dimensional case and the limit $U\to -\infty$.
For strong coupling limits spin models often turn out to be effective
Hamiltonians for models of correlated electrons. E.g., the antiferromagnetic
Heisenberg model is known to be the effective Hamiltonian for the Hubbard model
in the limit $U\to\infty$ at half-filling. In our case the effective
Hamiltonian is found to be the ferromagnetic Heisenberg chain.

We perturbate the supersymmetric Hubbard model by changing the
value of the nearest-neighbour Coulomb interaction $V$ in such a way that
the effective Hamiltonian in the limit of large attraction will be
a ferromagnetic non-critical $XXZ$ model ($\Delta \geq 1$,
see eq.\ (\ref{xxz}) below). In this regime the ground state is simply the
fully polarized ferromagnetic state $\ferroup$, and the spectrum has a gap
which vanishes in the limit $\Delta \rightarrow 1^{+}$.
The question of stability of the superconducting ground state of the
electronic model thus turns out to be closely related to the form of the
ground state of the ferromagnetic $XXZ$ model at fixed magnetization.

In the following we will be interested in the large-$U$ limit of generalized
Hubbard models with Hamiltonian
\ba
{\cal H} &=& -t\ \sum_{j=1}^L \sum_{\sigma =\up,\down}\left( \cd_{j\sigma}
             c_{{j+1},\sigma} + \cd_{{j+1},\sigma}c_{j\sigma} \right)
	  + V\sum_{j=1}^L (n_j-1)(n_{j+1}-1) \nonu
& &+X\sum_{j=1}^L\sum_{\sigma} ( \cd_{j\si} c_{{j+1},\si} + \cd_{{j+1},\si}
c_{j\si} ) (n_{j,-\si}+n_{{j+1},-\si}) \nonu
& &
 +Y\sum_{j=1}^L (\cd_{j\up} \cd_{j\down} c_{{j+1}\down} c_{{j+1}\up} +
 \cd_{{j+1}\up} \cd_{{j+1}\down} c_{j\down} c_{j\up}) \nonu
& &+ {J_{xy}\over 2}\sum_{j=1}^L \left(S_j^+S_{j+1}^- + S_j^-S_{j+1}^+
   \right)
   + J_z \sum_{j=1}^L S_j^z S_{j+1}^z \nonu
& &+ U \, \sum_{j=1}^L \bigl(n_{j\up}-\half)(n_{j\down}-\half\bigr)\ .
\label{EKS_Ham}
\ea
Here $\cd_{j\si}$ and $c_{j\si}$ are the usual electron
creation and annihilation operators, i.e.\ $\{ c_{j\si},\cd_{l\si'} \}
=\delta_{jl} \delta_{\si \si'}$, $n_{j\si} = \cd_{j\si}c_{j\si}$ is the
corresponding number operator and the components of the spin operators
are given by $S^z_j=\half \left(n_{j\up}-n_{j\down}\right)$,
$S^x_j = \half \left(S^+_j+S_j^-\right)$ and $S^y_j = {1\over 2i}
\left(S^+_j-S_j^-\right)$ with $S^+_j=\cd_{j\up}c_{j\down}$,
$S^-_j=\cd_{j\down}c_{j\up}$.

In \cite{dBKS} it has been investigated under which conditions the
Hamiltonian (\ref{EKS_Ham}) has an $\eta$-pairing ground state of
the form
\be
\grst = \left(\eta^\dagger\right)^N\vac
\label{eta}
\ee
with $\eta^\dagger = \sum_{j=1}^L\eta_j^\dagger = \sum_{j=1}^L
\cd_{j\down} \cd_{j\up}$. It has been found that this is true
for $t=X$, $2V=Y$,$V\leq 0$ and
\be
-U \geq 2\max \left(
V-\frac{J_z}{4},\ V+\frac{J_z}{4}+|\frac{J_{xy}}{2}|,\ 2V + 2|t|\right).
\label{bound}
\ee
Note that the first two conditions also guarantee the $\eta$-symmetry
of the Hamiltonian (\ref{EKS_Ham}) with $U=0$, i.e.
$[{\cal{H}}(U=0),\eta^\dagger]=0$.

For the supersymmetric Hubbard model of \cite{EKSa,EKSb} only
the Coulomb interaction $U$ is a free parameter. The other interaction
constants have fixed values: $X=t$, $V=-t/2$, $Y=-t$, and
$J_{xy}=J_z=2t$. The other integrable model (\ref{EKS_Ham}) has -- apart
from the Coulomb interaction $U$ -- only two non-vanishing interactions
constants, namely $t=X$ \cite{as}.

In the limit of strong attraction ($U\to -\infty$) in the Hamiltonian
(\ref{EKS_Ham}) one expects \footnote{This can be shown explicitly for
the exactly solvable cases \cite{EK,as}.} a gap for the single-particle
excitations  which is proportional to $|U|$. This means that only the motion
and interactions of the doubly occupied sites are important. Therefore all
the terms in (\ref{EKS_Ham}) can be dropped except for the pair-hopping term
$Y$ and nearest-neighbour Coulomb interaction $V$. Setting $Y=-1$ (since the
$\eta$-symmetry implies $Y\leq 0$) we then get the effective Hamiltonian
\be
{\cal H}_{eff} = -\sumnn \left( {\Delta\over 2}(n_j-1)(n_{j+1}-1)+
\cd_{j\up}\cd_{j\down}c_{j+1,\down}c_{j+1,\up}+
\cd_{j+1,\up}\cd_{j+1,\down}c_{j\down}c_{j\up}
\right)
\label{Heff1}
\ee
with $\Delta = 2V/Y=1$. Allowing for an interaction constant $\Delta \neq 1$
thus means a perturbation of the original model. In this letter we want
to investigate the effects of $\Delta > 1$ (i.e. $2V > Y$).
Note that this perturbation destroys the $\eta$-symmetry of the original
Hamiltonian.

In order to derive an effective spin Hamiltonian we first make a
partial particle-hole transformation $\cd_{j\up}\to \cd_{j\up}$ and
$\cd_{j\down}\to c_{j\down}$ on the $\down$-spins only.  This changes
$n_{j\up}\to n_{j\up}$ and $n_{j\down}\to 1-n_{j\down}$ and transforms
doubly-occupied sites into $\up$-spins and empty sites into
$\down$-spins. The transformed effective Hamiltonian (\ref{Heff1}) now
reads
\ba
{\cal H}_{eff} &=& -\sumnn \Bigl( {\Delta\over 2}(n_{j\up}-n_{j\down})
(n_{j+1,\up}-n_{j+1,\down})\nonu
& &\qquad\quad\qquad +\cd_{j\up}c_{j\down}\cd_{j+1,\down}c_{j+1,\up}+
\cd_{j\down}c_{j\up}\cd_{j+1,\up}c_{j+1,\down}\Bigr).
\label{Heff2}
\ea
Using the spin operators introduced above and
$S_j^\alpha =\half \sigma_j^\alpha$ (where $\sigma_j^\alpha$ are the
Pauli matrices, $\alpha =x,y,z$) the Hamiltonian (\ref{Heff2}) becomes
\be
{\cal H}_{eff} = -\half\sumnn \left(\sigma_j^x\sigma_{j+1}^x +
\sigma_j^y \sigma_{j+1}^y+ \Delta \sigma_j^z \sigma_{j+1}^z \right)
\label{xxz}
\ee

Under the particle-hole transformation -- which maps the fermionic vacuum
$\vac$ onto the ferromagnetic state $\ferro = \prod_{j=1}^L\cd_{j\down}
\vac$ -- the $\eta$-pairing state (\ref{eta}) translates into
\be
\grst = \left( S^+\right)^N\ferro
\label{grst1}
\ee
which is the ground state of the isotropic Heisenberg ferromagnet in
the sector with fixed magnetization $S^z=N-{L\over 2}$.

In the following we will determine how the form (\ref{grst1}) of the
ground state changes for $\Delta >1$ by proving a conjecture by
Gaudin \cite{GAUDIN}.

Hamiltonian (\ref{xxz}) has long been known to be solvable by means of
Bethe Ansatz \cite{B,YY}. In the sector with a fixed number $N$ of upturned
spins, the eigenvalues are determined by the equations \cite{YY,G,J}
\be
\left(\frac{\sin (\lambda_{j}+\frac{i\gamma}{2})}{\sin (\lambda_{j}-
\frac{i\gamma}{2})}\right)^{L}=-\prod_{l=1}^{N}\frac{\sin (\lambda_{j}-
\lambda_{l}+i\gamma)}{\sin (\lambda_{j}-\lambda_{l}-i\gamma)}\ ,
\label{bae2}
\ee
\be
E=-\frac{L\Delta}{2}+2\sum_{j=1}^{N}\frac{\sinh ^{2}\gamma}{\cosh
\gamma-\cos 2\lambda_{j}}\ .
\label{en2}
\ee
We will consider all possible types of string solutions
%\begin{displaymath}
\be
\lambda^{(n)}_{\alpha ,k}=\lambda^{(n)}_{\alpha}-\frac{i\gamma}{2}
(n+1-2k) \qquad (k=1,2, \ldots ,n \ \mbox{with} \ n\in Z_{+}).
%\end{displaymath}
\label{string}
\ee
Periodicity along the real axis allows one to take the real center
of the string, $\lambda^{(n)}_{\alpha}$, in the restricted range
$(-\pi /2,\pi /2]$. As usual, it is possible to reduce (\ref{bae2})
to a set of equations for the string centers. Choosing %\begin{displaymath}
\be
\phi(\lambda,\alpha) \doteq
i\log \left(-\frac{\sin (\lambda+i\alpha)}{\sin (\lambda-i\alpha)}\right)
=2\arctan (\tan \lambda \coth \alpha)
%\end{displaymath}
\label{phase}
\ee
we find
\be
\frac{1}{2\pi}t_{n}(\lambda^{(n)}_{\alpha})-\frac{1}{2\pi L}
\left[ \sum_{m \neq n}\sum^{M_{m}}_{\beta=1}\Theta_{n,m}
(\lambda^{(n)}_{\alpha}-\lambda^{(m)}_{\beta})+
\sum^{M_{n}}_{\beta =1}\Theta_{n,n}(\lambda^{(n)}_{\alpha}-
\lambda^{(n)}_{\beta})\right]=\frac{I^{(n)}_{\alpha}}{L}
\label{bae3}
\ee
where $M_{m}$ denotes the number of $m$-strings, $I^{(n)}_{\alpha}$
is integer (half-odd) if $L+M_{n}+1$ is even (odd) and
\begin{eqnarray}
t_{n}(\lambda)&=&\phi\left(\lambda,\frac{n\gamma}{2}\right) \nonu
\Theta_{n,m}(\lambda)&=&\phi\left(\lambda,\frac{\gamma}{2}(n+m)\right)+
\phi\left(\lambda,\frac{\gamma}{2}|n-m|\right)+\sum^{\min(m,n)-1}_{k=1}
2\phi(\lambda,\frac{\gamma}{2}(n+m-2k)) \nonu
\Theta_{n,n}(\lambda)&=&\phi(\lambda,n\gamma)+\sum^{n-1}_{k=1}
2\phi(\lambda,k\gamma)
\label{stringphase}
\end{eqnarray}
The energy of a $n$-string, derived from (\ref{en2}), is
\be
e_{n}(\lambda)=2\sinh\gamma \ \frac{\sinh n\gamma}{\cosh n\gamma-\cos
2\lambda}
\label{enstr}
\ee
and therefore the total energy, neglecting the immaterial additive
constant, is
\be
E=2\sinh \gamma \ \sum^{\infty}_{n=1}\sum^{M_{n}}_{\beta=1}
\frac{\sinh n\gamma}{\cosh n\gamma -\cos 2\lambda^{(n)}_{\beta}}
\ee
where $\{\lambda^{(n)}_{\beta}\}$ are the real string centers.
We are interested in the ground state in a sector of fixed
magnetization where
\be
\sum^{\infty}_{n=1}nM_{n}=N, \qquad \lim_{L \rightarrow \infty}
\frac{N}{L}=a, \qquad  0<a \leq \frac{1}{2} .
\label{mag}
\ee
In principle one should find all solutions of the Bethe Ansatz equations
for the string centers $\lambda^{(n)}_{\alpha}$ which are compatible
with (\ref{mag}) and compare the respective energies. Yet, for
a string of length $n$, lower and upper bounds of the energy
(\ref{enstr}) are given by
\be
2\sinh \gamma \frac{\sinh n\gamma}{\cosh n\gamma +1} \leq
2\sinh \gamma \frac{\sinh n\gamma}{\cosh n\gamma -\cos 2\lambda} \leq
2\sinh \gamma \frac{\sinh n\gamma}{\cosh n\gamma -1}
\ee
for $\lambda=\pi/2$ and $\lambda=0$ respectively. We conclude that
the ground state must contain exactly one string of length $N$ if we
can show that, for any other string configuration $(M_{1},M_{2},
\ldots M_{N-1},0)$ with $\sum^{N-1}_{n=1}nM_{n}=N$, we have
\be
\frac{\sinh N\gamma}{\cosh N\gamma -1} < \sum^{N-1}_{n=1}M_{n}
\frac{\sinh n\gamma}{\cosh n\gamma +1}
\label{ineq}
\ee
First we observe that, in the sector under consideration, any
configuration $(M_{1},M_{2}, \ldots, M_{N-1},0)$ can be obtained
starting from one of the configurations where two strings only
are present, $M_{n}=1$, $M_{m}=1$, $n+m=N$, and performing steps
in which a string is broken into two smaller strings
%\begin{displaymath}
\ba
& &(M_{1},\ldots, M_{n_{1}}, \ldots, M_{n_{2}}, \ldots, M_{n}, \ldots )
\nonu
& &\qquad\rightarrow  (M_{1},\ldots, M_{n_{1}}+1,\ldots, M_{n_{2}}+1,\ldots,
M_{n}-1,\ldots )
%\end{displaymath}
\label{brakestring}
\ea
with $n_{1}+n_{2}=n$. In each step the RHS of (\ref{ineq}) increases since
%\begin{displaymath}
\be
\frac{\sinh n_{1}\gamma}{\cosh n_{1}\gamma +1}+ \frac{\sinh n_{2}\gamma}
{\cosh n_{2}\gamma +1} \geq \frac{\sinh n\gamma}{\cosh n\gamma +1}
\qquad  (n_{1}+n_{2}=n).
%\end{displaymath}
\ee
This inequality is guaranteed by the fact that $f(0)=0$ and
$f''(x)<0$ in $(0,+\infty)$, where $f(x)=\frac{\sinh \gamma x}
{\cosh \gamma x +1}$, and consequently $f(x_{1})+f(x_{2}) \geq
f(x_{1}+x_{2})$. We conclude that, to demonstrate (\ref{ineq})
it is sufficient to consider configurations on the RHS made up of
two strings only. Again concavity of $f(x)$ shows that it is enough
to have
%\begin{displaymath}
\be
\frac{\sinh N\gamma}{\cosh N\gamma -1} < \frac{\sinh \gamma}
{\cosh \gamma +1}+\frac{\sinh (N-1)\gamma}{\cosh (N-1)\gamma+1}
%\end{displaymath}
\label{ineq2}
\ee
and, for fixed $\gamma$, this is certainly the case for $N$ large
enough\footnote{Note that for large $N$ the difference
$\frac{\sinh (N-1)\gamma}{\cosh (N-1)\gamma+1} -
\frac{\sinh N\gamma}{\cosh N\gamma -1}$ is of order $e^{-N\gamma}$.}.
This concludes the proof that the ground state is given by
a solution of the Bethe Ansatz equations  with only one $N$-string
confirming a result predicted by Gaudin \cite{GAUDIN}. For the isotropic
case $\Delta =1$ a similar result was already obtained by Bethe \cite{B}.

The string center $\lambda^{(N)}$ of the Bethe state with a single $N$-string
(\ref{string}) can easily be determined from (\ref{bae3}).
On a finite lattice there are $L$ solutions
%\begin{eqnarray*}
\ba
&&\lambda^{(N)}=\arctan \left(\tanh \frac{N\gamma}{2} \tan
\frac{\pi I^{(N)}} {L}\right)\ ,\nonu
&&I^{(N)}= -\frac{L}{2}+1, -\frac{L}{2}+2, \ldots, \frac{L}{2}
%\end{eqnarray*}
\ea
(this holds for both $L$ even and $L$ odd) with the ground state at
$I^{(N)}=\frac{L}{2}$ and energy given by (\ref{enstr}). The energies
of all these states become degenerate in the limit $L \rightarrow
\infty$ (and consequently $N \rightarrow \infty$) with limiting
value $2\sinh \gamma$ and energy differences vanishing like
$O(e^{-N\gamma})$. Consequently we have an infinite degeneracy of the
ground state.

Low-lying excitations are given by configurations where the
$N$-string is broken into a finite number of shorter strings.
In this situation, the "interaction term" in equation (\ref{bae3})
for the centers $\lambda_\alpha^{(n)}$ goes to
zero like $1/L$. Therefore, the energy of each string coincides with
the bare energy (\ref{enstr}) and no dressing needs to be considered.
In the most general excited state configuration $(M_{1},M_{2},\ldots,
M_{N-1})$ with $\sum^{N-1}_{n=1}nM_{n}=N$, the total number of strings
$\sum^{N-1}_{n=1}M_{n}$ remains finite and the length of one or more
strings has to diverge. Each of these diverging length strings
contributes $2\sinh \gamma$ to the energy, regardless of their position,
while the finite length ones have energy (\ref{enstr}). It is easily
seen that the spectrum has a gap
\be
\Delta E=2(\cosh \gamma -1)
\ee
obtained by taking $M_{1}=1$, $M_{N-1}=1$.

The above results show that the ground state of the ferromagnetic
$XXZ$ chain (\ref{xxz}) with $\Delta > 1$ at fixed magnetization
$S^z =N-\frac{L}{2}$ is a bound state of the form
\be
\grst = \sum_{\{x_1,\ldots,x_N\}} \psi(x_1,\ldots,x_N)
\prod_{j=1}^N\sigma^+_{x_j}\ferro .
\label{bstate1}
\ee
Here $\psi(x_1,\ldots,x_N)$ is a bound state wave function, i.e.\
it decays exponentially with respect to all coordinate differences
$|x_j - x_l| \to \infty$ ($j,l = 1,\ldots, N$ with $j \neq l$) \cite{GAUDIN}.
The wave function $\psi(x_1,\ldots,x_{N})$ for $x_1 < x_2 < \ldots < x_{N}$
is given by the Bethe-Ansatz expression
\be
\psi(x_1,\ldots,x_{N}) = \sum_{P\in S_N}\prod_{j=1}^{N}\left[\frac
{\sin (\lambda_{P(j)}-i\frac{\gamma}{2})}{\sin (\lambda_{P(j)}+i
\frac{\gamma}{2})}\right]^{x_{j}}\prod_{\stackrel{j<l}{P(j)>P(l)}}
\frac{\sin (\lambda_{P(l)}-\lambda_{P(j)}+i\gamma)}{\sin (\lambda_{P(l)}
-\lambda_{P(j)}-i\gamma)}
\label{wave1}
\ee
where $S_N$ denotes the permutation group. We now specialize to the state
containing exactly one $N$-string. It can be seen easily
that only the term where $P$ is the identity has a non-vanishing
contribution. Introducing the new variables $z_0 = {1\over N}\sum_{j=1}^N
x_j$ and $z_j=x_{j+1}-x_j$ ($j=1,\ldots,N-1$), i.e.\ $x_j=z_0+{1\over N}
\sum_{k=1}^{N-1}kz_k - \sum_{k=j}^{N-1}z_k$
we finally can rewrite the wave function as
\be
\psi(x_1,\ldots,x_{N}) =
C(\lambda,z_0,N)
\prod_{l=1}^{N-1}\left[ \left(\frac{\sin(\lambda-\frac{i\gamma}{2}(N-2l))}
{\sin(\lambda-\frac{i\gamma}{2}N)}\right)
\left(\frac{\sin(\lambda-\frac{i\gamma}{2}N)}{\sin(\lambda+\frac{i\gamma}{2}N
)}\right)^{l/N}\right]^{z_l}
\label{wave3}
\ee
where the constant $C(\lambda,z_0,N)$ depends only on $\lambda$, $z_0$ and
$N$. The form (\ref{wave3}) shows directly that the wave function describes
a bound state since it decays exponentially as function of the
coordinate differences $z_j=x_{j+1}-x_j$.

For the fermionic model the ground state is then of the form
\be
\grst = \sum_{\{x_1,\ldots,x_N\}} \psi(x_1,\ldots,x_N)
\prod_{j=1}^N(\cd_{x_j\down}\cd_{x_j\up})\vac \ .
\label{bstate}
\ee
This state has no ODLRO, i.e.\ ${\langle \psi_N | \cd_{j\down} \cd_{j\up}
c_{l\up} c_{l\down} | \psi_N \rangle \over \langle \psi_N |
\psi_N \rangle}\quad \stackrel{|l-j| \to \infty
}{\longrightarrow} \quad 0$. This can be seen as follows:

Using the $SU(2)$ commutation relations for $\eta_j = c_{j\up}c_{j\down}$
\cite{EKSa} it is easy to see that
\be
{\langle \psi_N | \eta_{j}^\dagger \eta_{l} | \psi_N \rangle
\over \langle \psi_N |\psi_N \rangle}
= {\cal N}\sum_{\{x\}}\psi^*(x_1,\ldots,x_{N-1},j)\psi(x_1,\ldots,x_{N-1},l)
\label{odlro1}
\ee
where ${\cal N}$ is a normalization constant. Since $\psi(x_1,\ldots,x_N)$
decays exponentially as function of all coordinate differences a significant
contribution to the sum in (\ref{odlro1}) comes only from $x_1\approx x_2
\approx \ldots \approx x_{N-1}\approx j$ and $x_1\approx x_2
\approx \ldots \approx x_{N-1}\approx l$. This means that for
$|j-l|\to\infty$ at least one of the factors in each term and thus
the whole sum decays exponentially.

This argument shows that superconductivity in the one-dimensional
supersymmetric Hubbard model is destroyed by a perturbation with $\Delta
> 1$. Instead, the form of the ground state suggests a tendency to
phase separation. This is not surprising since numerical investigations
have shown that superconducting phases in electronic models appear
quite generally in the vicinity of phase separation \cite{dagotto}.
In higher dimensions however the situation is somewhat different. By
analogy with the non-ideal Bose gas and preliminary results from
perturbation theory \cite{kareljan} we expect the superconducting
ground state to be stable under the kind of perturbation considered here.

\section*{Acknowledgement}
VEK is supported in part by NSF grant No.~PHY-9321165. GA was
partially supported by the NSF under grant \# PHY9309888.
AS\ thanks the Deutsche Forschungsgemeinschaft for financial support.
We thank Prof.\ J.H.H.\ Perk for drawing our attention to reference
\cite{GAUDIN} and Prof.~M.~Gaudin for clarifying comments about \cite{GAUDIN}.

\end{document}